\documentclass[aps,prl,twocolumn]{revtex4-1}
\usepackage{bm}
\usepackage{graphicx}
\usepackage{amsmath}
\usepackage{amssymb}
\usepackage[utf8]{inputenc}
\usepackage[T1]{fontenc}
\usepackage{color}
\usepackage{upgreek}
\usepackage{subfigure}
\usepackage[normalem]{ulem}
\usepackage[unicode=true,colorlinks=true,citecolor=blue]{hyperref}
\usepackage{placeins}
\usepackage{lipsum}
\usepackage{epsfig}
\usepackage[utf8]{inputenc}
\usepackage{makecell}
\usepackage{wrapfig}
\newcommand{\nix}[1]{}
\definecolor{greenI}{rgb}{0, .4, 0}

\usepackage[version=4]{mhchem}
\usepackage{glossaries}
\setacronymstyle{long-short}
\glsdisablehyper
\usepackage{siunitx}

\newcommand{\kVcm}[1]{\SI{#1}{\kilo\volt\per\centi\meter\squared}}
\newcommand{\kWcm}[1]{\SI{#1}{\kilo\watt\per\centi\meter\squared}}
\newcommand{\MWcm}[1]{\SI{#1}{\mega\watt\per\centi\meter\squared}}

\newcommand{\MilliElectronVolt}[1]{\SI{#1}{\milli\electronvolt}}

\newcommand{\NanoMeter}[1]{\SI{#1}{\nano\meter}}
\newcommand{\NanoAmpere}[1]{\SI{#1}{\nano\ampere}}

\newcommand{\TeraHertz}[1]{\SI{#1}{\tera\hertz}}
\newcommand{\NanoSecond}[1]{\SI{#1}{\nano\second}}

\newcommand{\Degree}[1]{\SI{#1}{\degree}}

\newacronym{lpge}{LPGE}{linear photogalvanic effect}
\newacronym{cpge}{CPGE}{circular photogalvanic effect}
\newacronym{cpde}{CPDE}{circular photon drag effect}
\newacronym{lpde}{LPDE}{linear photon drag effect}
\newacronym{tblg}{tBLG}{twisted bilayer graphene}
\newacronym{hbn}{hBN}{hexagonal boron nitride}
\newacronym{pge}{PGE}{photogalvanic effect}
\newacronym{pde}{PDE}{photon drag effect}
\newacronym{cnp}{CNP}{charge neutrality point}
\newacronym{thz}{THz}{terahertz}

\begin{document}

\title{Nonlinear intensity dependence of photogalvanics and photoconductance induced by terahertz laser radiation in twisted bilayer graphene close to magic angle}

\author{S. Hubmann$^1$, P. Soul$^1$, G. Di Battista$^2$, M. Hild$^1$, K. Watanabe$^3$, T. Taniguchi$^4$, D.K. Efetov$^2$, and S.D. Ganichev$^{1,5}$}
%
\affiliation{$^1$Terahertz Center, University of Regensburg, 93040 Regensburg, Germany}

\affiliation{$^2$ICFO - Institut de Ciencies Fotoniques, The Barcelona Institute of Science and Technology, Castelldefels, Barcelona 08860, Spain}

\affiliation{$^3$Research Center for Functional Materials, 
	National Institute for Materials Science, 1-1 Namiki, Tsukuba 305-0044, Japan}

\affiliation{$^4$International Center for Materials Nanoarchitectonics, National Institute for Materials Science, 1-1 Namiki, Tsukuba 305-0044, Japan}
	
\affiliation{$^5$CENTERA Laboratories, Institute of High Pressure Physics, Polish Academy of Sciences PL-01-142 Warsaw, Poland}

\begin{abstract}
We report on the observation of the nonlinear intensity dependence of the bulk photogalvanic current and photoconductivity in the twisted graphene with small twist angles close to the second magical angle. We show that terahertz radiation results in the photoresponses, which is caused by indirect optical transitions (free carrier absorption), direct interband transitions and optical transitions between Moiré subbands. The relative contribution of these absorption channels depends on the Fermi level position with respect to the multiple  Moiré subbands of the twisted graphene. The interplay of these absorption channels results in oscillations of the photoresponses with variation of the gate voltage. We show that the photoresponse saturates at high intensities. For different absorption channels it has different intensity dependencies and saturation intensities. The latter depends non-monotonically on the Fermi level position, which is controlled by the gate voltage. 
\end{abstract}

\maketitle

\section{Introduction}
\label{introduction}
In the last decade, \gls{tblg} has been shown to be a promising and versatile material for fundamental research and applications. Several exotic states as tunable superconductivity and correlated insulators \cite{Cao2018a,Cao2018,Wu2018,Po2018,Xu2018a,Yankowitz2019,Lu2019,Shen2020,Liu2020}, emergent ferromagnetism \cite{Sharpe2019}, van-Hove singularities \cite{Li2009,Yan2012a}, and twist-controlled tunneling \cite{Mishchenko2014} have been detected in \gls{tblg}. The material system is composed by two graphene sheets which are rotated with respect to one another by a specific twist angle $\theta$. Consequently, a Moiré superlattice is formed, and the value of the twist angle determines the size of the Moiré unit cell, having crucial impact on the properties of the material system, for further references see e.g.~ \cite{Santos2007,Morell2010,Mele2010,Bistritzer2011,Luican2011,Rozhkov2016,Kim2017a,Koshino2018,RibeiroPalau2018,Yoo2019,Kerelsky2019,Jiang2019,Serlin2020,Xie2020,Stepanov2020,Cao2020,Lu2021,Bernevig2021,Cao2021,Hao2021,Das2021,Wu2021}. A special case arises if the twist angle $\theta$ is lower than \Degree{1}, where an extraordinarily large Moiré supercell is formed. In this case a highly complex energy spectrum consisting of several  minibands is formed \cite{Morell2010,Choi2019,Otteneder2020,Lu2021,Wu2021}. Such low-angle \gls{tblg} devices are in focus of the current research revealing some of the fascinating properties of this material system. However, there are only few works on optoelectronic phenomena in \gls{tblg} \cite{Xin2016,Yin2016a,Otteneder2020,Sunku2020,Hesp2021,Sunku2021}. Most recently, we reported on the \glspl{pge} excited by low-intensity \gls{thz} laser radiation in bulk and edges of \gls{tblg} \cite{Otteneder2020}. This work demonstrated that \gls{thz} radiation results in a redistribution of carriers in momentum space caused by the asymmetric scattering in gyrotropic \gls{tblg} and provided evidence for the multitude of  bands as well as the reduction of the symmetry of small-angle \gls{tblg}. 

Here we report on the observation of the complex nonlinear intensity dependence of the \gls{thz} radiation-induced \gls{pge}
in small-angle \gls{tblg}. 
We show that the overall behavior of the photocurrent is crucially influenced by the applied back gate voltage $U_{\text G}$. The photocurrent exhibits pronounced oscillations with the variation of $U_{\text G}$ and its nonlinear behavior changes qualitatively: Depending on the back gate voltage the photocurrent shows either linear behavior, saturate at high intensities, or even exhibits a change of sign with increasing radiation intensity.
Furthermore, we observed the \gls{thz} photoconductance, which shows sign-alternating oscillations with $U_{\text G}$ and also behaves nonlinearly with increasing radiation intensity.
The observed oscillations are related to the resistance oscillations and are shown to be caused by the shifting of the Fermi energy across the multitude of Moiré bands.
The analysis demonstrates that the complex behavior of the photoresponse with increasing intensity is due to the interplay between two mechanisms of radiation absorption: Indirect optical transitions (Drude-like) give a contribution to the photoresponse as well as direct optical transitions between the Moiré minibands. 
The obtained characteristics and the functional behavior of \gls{thz} radiation saturation demonstrate that \gls{tblg} can be used for the development of \gls{thz} mode-locking laser systems, which are already used in the infrared/visible spectral ranges using monolayer graphene saturable absorbers.

\section{Samples and methods}
\label{samples_methods}

A ''cut and stack'' method was used to assemble the hBN/\gls{tblg}/hBN/graphite heterostructure~\cite{Kim2017a}. The graphene planes were first rotated by a target angle of \Degree{1} and then the \gls{tblg} relaxed into the final twist angle of $\theta \approx \Degree{0.3}$.The stack was then etched into Hall bar geometry ($\SI{4}{\micro\meter} \times \SI{19}{\micro\meter}$), see inset in Fig.~\ref{Fig1}, and one-dimensional contacts were made evaporating \ce{Cr}/\ce{Au} (5/\NanoMeter{50}) electrodes \cite{Wang2013}. We define the directions along the long side of the Hall bar as $x$-direction, and perpendicular to that as $y$-direction. The \gls{tblg} was encapsulated between hBN layers and a graphite layer on the bottom was used as a local back gate.

Transport measurements were carried out in a He-4 based variable temperature insert cryostat in a temperature range from liquid helium to room temperature. We used standard low-frequency lock-in technique to measure the resistance in two-terminal geometry with a \NanoAmpere{100} excitation  current at a frequency of \SI{12}{Hz}. 
To continuously tune the Fermi energy in our system we apply a gate voltage $U_\text{G}$ in the range of $\pm \SI{4}{\volt}$ to the graphite back gate. Note that for different sample cooldowns, the \gls{cnp} can occur at slightly different gate voltages. This is caused by, e.g., cooldown-dependent charge trapping in the insulator ~\cite{Candussio2021,Dantscher2017}. In order to account for this, below we use the effective gate voltage $U_{\text{G,eff}}= U_{\text{G}}-U_{\text{CNP}}$.

The \gls{thz} electric field has been applied using an line-tunable, pulsed high-power molecular gas laser
\cite{Shalygin2007,Plank2016}, pumped by transversally excited atmospheric pressure (TEA) CO$_2$-laser~\cite{Ganichev2003}.
In this study we used frequency lines of 0.6, 0.8, 1.1, 2, and \TeraHertz{3.3} (photon energies of 2.5, 3.2, 4.4, 8.3, and \MilliElectronVolt{13.8}) with intensities up to few \si{\mega\watt\per\centi\meter\squared}, a pulse duration of about \NanoSecond{100} and a repetition rate of \SI{1}{\hertz}. The Gaussian shaped laser beam was focused onto the sample using an parabolic mirror yielding a spot-size of about \SI{1.5}{\milli\meter}. Note that the size of the laser spot is much larger than the sample dimension, which ensures uniform illumination of the sample.
The sample temperature has been controllably varied by using an optical continuous flow cryostat, which allowed to tune the temperature between liquid helium temperature and room temperature.
The radiation intensity was modified using a crossed polarizer setup \cite{Hubmann2019,Candussio2021a} and the radiation electric field was varied using half- and quarter-wave plates, see Ref.~\cite{Belkov2005}. The rotation of the radiation electric field $\bm E$ with respect to the $y$-direction is described by the azimuth angle $\alpha$, see inset in Fig.~\ref{Fig1}. 
All experiments were conducted with laser beam illuminating the sample at normal incidence. 

The photocurrent signals were measured as a voltage drop along a load resistor $R_{\text L}=\SI{50}{\ohm}$, while the sample was unbiased. For the measurements of the photoconductivity the sample was additionally biased by a dc bias voltage $U_{\text{dc}}= \pm\SI{0.1}{\volt}$. In this kind of measurements, the signal consists of two contributions caused by the generation of the photocurrent and the change of conductivity upon irradiation. The former one is independent of the bias polarity, whereas the latter one changes its sign upon reversing $ U_{\text{dc}}$. Using this fact, we extracted the photoconductivity signal from the photocurrent signals by subtracting the voltage drops obtained for positive and negative bias voltages and dividing by 2.

\section{Results}
\label{results}
\begin{figure}
	\centering
	\includegraphics[width=\linewidth]{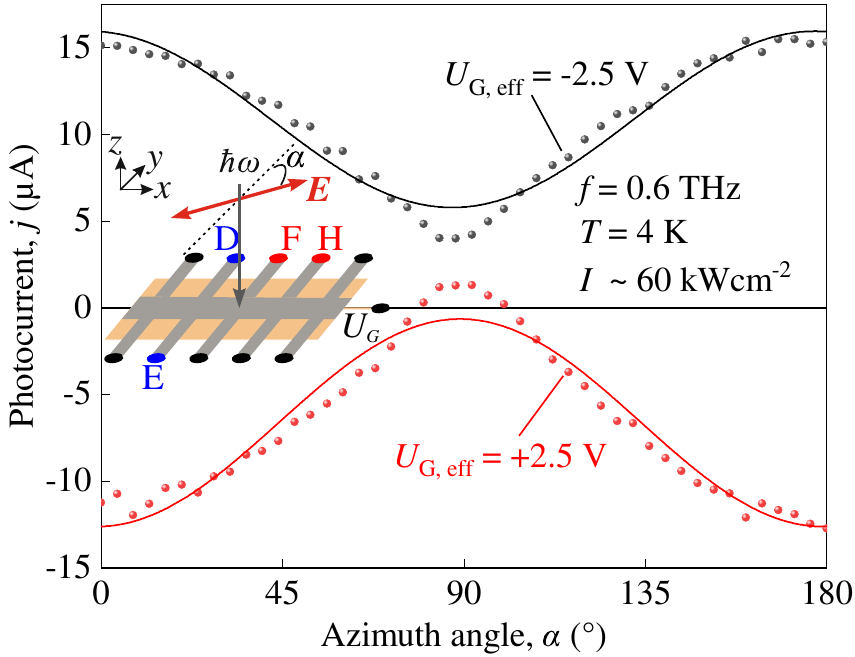}
	\caption{Photocurrent as a function of the radiation electric field orientation defined by the angle $\alpha$ at fixed back gate voltages of $U_{\text {G,eff}}=\SI{\pm2.5}{\volt}$ and a radiation frequency of \TeraHertz{0.6}. The inset shows a sketch of the sample and defines the azimuth angle $\alpha$. The data are fitted by Eq.~\eqref{parameters_definition} with fitting parameters $j_{0}$, $j_{1}$, and $j_{2}$.}
	\label{Fig1}
\end{figure}

The illumination of the \gls{tblg} structure with linearly polarized \gls{thz} radiation resulted in the emergence of a photocurrent measured across the sample (in $y$-direction, see inset in Fig.~\ref{Fig1}). The photocurrent is detected in the whole temperature range studied here.
Amplitude and sign of the photocurrent depend on the orientation of the radiation electric field vector $\bm E$ and the value of the gate voltage. The polarization dependencies are exemplary shown in Fig.~\ref{Fig1} for the contact pair D-E, a radiation frequency of \TeraHertz{0.6} and two complementary gate voltages.

The dependence of the photocurrent on the azimuth angle $\alpha$ defining the orientation of $\bm E$ with respect to the $y$-axis, is described by \cite{Otteneder2020}
\begin{equation}
	j_y=j_{0}-j_{1}\cos(2\alpha)-j_{2}\sin(2\alpha)\,.
	\label{parameters_definition}
\end{equation}
This corresponds to a linear combination of the Stokes parameter $P_0$, $P_{\text L1} = \cos 2\alpha$ and $P_{\text L2} = \sin 2\alpha$ weighted by the coefficients $j_{0}$, $j_{1}$ and $j_{2}$. Note that the contribution proportional to $P_{\text L2} = \sin 2\alpha$ is vanishingly small. Applying elliptically (circularly) polarization in the $\lambda/4$-plate setup we also detected a circular photocurrent contribution $j_{\text{circ}}$, which is proportional to the radiation helicity and reverses its sign by switching between right- and left-handed circularly polarized radiation. As addressed above, the photocurrent was observed across the sample only and was vanishingly small in measurements along the sample edges, e.g., along the contacts F-H, see inset in Fig.~\ref{Fig1}. The overall behavior of the photocurrent reveals that it originates in the bulk \gls{pge} previously observed in Ref.~\cite{Otteneder2020}, and is caused by the asymmetric scattering of photoexcited carriers in low-symmetric \gls{tblg} structures. The detailed discussion of the photocurrent formation is presented in Ref.~\cite{Otteneder2020} and, consequently, will not be discussed here. 

While the experiments in Ref.~\cite{Otteneder2020} were carried out applying radiation of the order of fractions of \si{\watt\per\centi\meter\squared}, in the present measurements we used substantially higher (by about six orders of magnitude) intensities in order of tens of \si{\kilo\watt\per\centi\meter\squared}.
As we show below this drastic increase in radiation intensity results in a nonlinear behavior of the photocurrent.

\begin{figure}
	\centering
	\includegraphics[width=\linewidth]{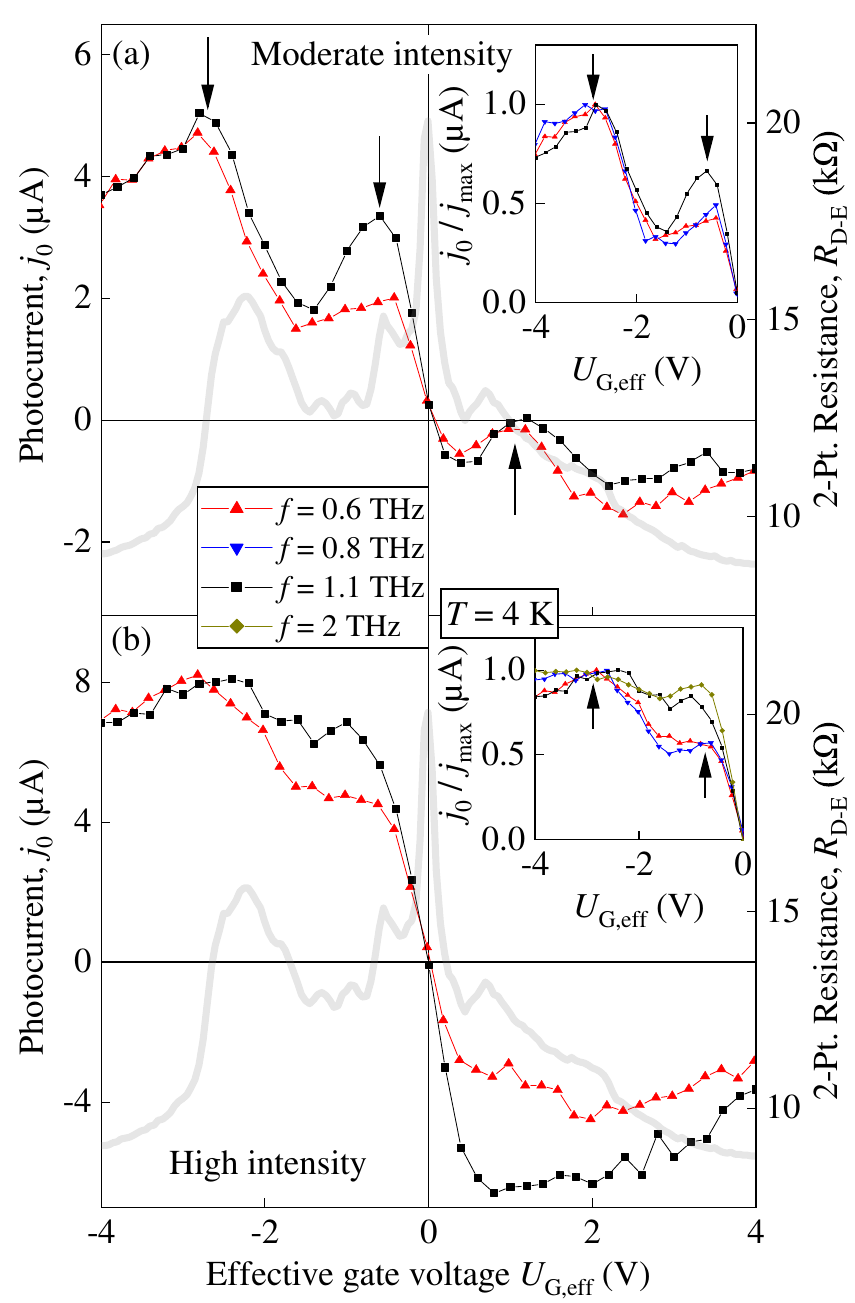}
	\caption{Dependences of the polarization-independent photocurrent contribution (s. Eq.~\eqref{parameters_definition}) for the $y$-direction (see inset in Fig.~\ref{Fig1}) on the effective back gate voltage $U_{\text{G,eff}}$ for different radiation frequencies. The thick gray lines show the corresponding two-point resistance $R_{\text{D-E}}$. The insets show a zoom-in of the corresponding dependencies for negative gate voltages and more radiation frequencies. Note that the data in the insets are normalized to their respective maximum value. The corresponding radiation intensities in Panel (a) are: \kWcm{15} (\TeraHertz{0.6}), \kWcm{15} (\TeraHertz{0.8}), \kWcm{20} (\TeraHertz{1.1}), 
	The radiation intensities in Panel (b) are: \kWcm{60} (\TeraHertz{0.6}), \kWcm{170} (\TeraHertz{0.8}), \kWcm{190} (\TeraHertz{1.1}), and \kWcm{1100} (\TeraHertz{2}) 
	. Note that the corresponding radiation intensities in (b) are by orders of magnitude higher  than in (a).}
	\label{Fig2}
\end{figure}

Varying the voltage applied to the gate we observed oscillations of the photocurrent with $U_{\text{G,eff}}$ at moderate radiation intensities of about $\kWcm{15}$. These are observed at low temperature only and are shown exemplary for the polarization-independent photocurrent contribution $j_{0}$ in Fig.~\ref{Fig2}(a).  The oscillation maxima are indicated by vertical arrows. The data are shown for different radiation frequencies, see Fig.~\ref{Fig2}(a) and the inset. The figure also shows the gate voltage dependence of the two-point resistance (gray line), demonstrating that the observed photocurrent oscillations correlates with those of the sample resistance.
The photocurrent oscillations are caused by transitions between the multitude of Moiré subbands and have the same origin as those observed at substantially lower intensity level ($\approx\si{\watt\per\centi\meter\squared}$)in Ref.~\cite{Otteneder2020}. The almost flat dispersion of the subbands results in sharp changes of the resistance with the variation of the Fermi energy resulting in the filling and emptying of these bands. This process enables/disables direct optical transitions and modifies the Drude-like intraband absorption. Consequently, one obtains the oscillations of the photocurrent. Note that the photocurrent changes its sign in the vicinity of the \gls{cnp}, because of the sign change of the carrier charge.


Strikingly, a substantial increase of the radiation intensity doesn't change the photocurrent magnitude much. This demonstrates the nonlinear behavior of the photocurrent. Furthermore, the increase in radiation intensity modifies the gate voltage dependence, see Fig.~\ref{Fig2}(b). While the oscillations are still resolved, they become less pronounced. This reveals that the nonlinearity is different for different gate voltages. A similar behavior was observed for all frequencies, see insets in Fig.~\ref{Fig2}. Note that the highest intensities applied scale from about \kWcm{50} to \MWcm{1} depending on the radiation frequency, which is caused by the line-dependent laser power limit. 


\begin{figure}
	\centering
	\includegraphics[width=\linewidth]{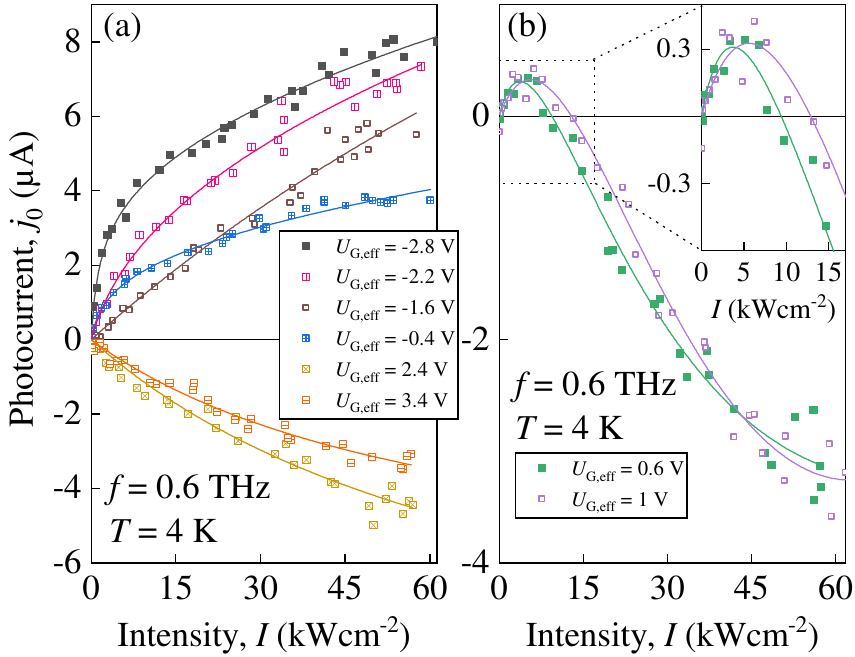}
	\caption{Dependences of the polarization-insensitive photocurrent contribution in the $y$-direction (see inset in Fig.~\ref{Fig1}) on the radiation intensity $I$ at a radiation frequency of \TeraHertz{0.6} for different effective gate voltages. The inset in panel (b) shows a zoom-in for low intensities as emphasized by the dashed rectangle. The curves show fits according to Eq.~\eqref{eq_photocurrent_fit}. The corresponding fit parameters are given in Tab.~\ref{tab1}.}
	\label{Fig6}
\end{figure}

\begin{figure}
	\centering
	\includegraphics[width=\linewidth]{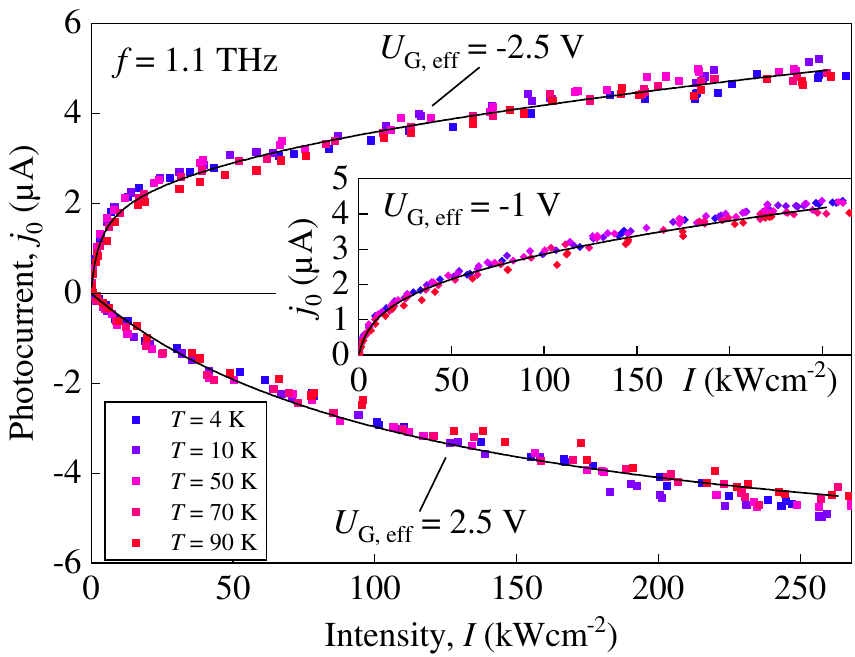}
	\caption{Dependences of the polarization-independent photocurrent contribution in the $y$-direction (see inset in Fig.~\ref{Fig1}) on the radiation intensity for different gate voltages and various temperatures. The curves show fits according to Eq.~\eqref{eq_photocurrent_fit}. The corresponding fit parameters are given in Tab.~\ref{tab1}.}
	\label{Fig8}
\end{figure}

In order to study the nonlinearity we investigated the intensity dependence of the photocurrent. Figure~\ref{Fig6} shows the corresponding dependencies of the photocurrent obtained at several gate voltages. The figure demonstrates the 
complex behavior of the photocurrent with increase of the radiation intensity and reveals that the nonlinearity type depends on the gate voltage. While for most gate voltage the current saturates with rising intensity, see Fig \ref{Fig6}(a), for low positive gate voltages close to the oscillation extrema the photocurrent exhibits a change of sign with increasing intensity, see Fig. \ref{Fig6}(b). Figure~\ref{Fig8} shows that at an increase of temperature up to about $\SI{100}{\kelvin}$, the photocurrent changes neither the magnitude nor the nonlinearity.

\begin{figure}
	\centering
	\includegraphics[width=\linewidth]{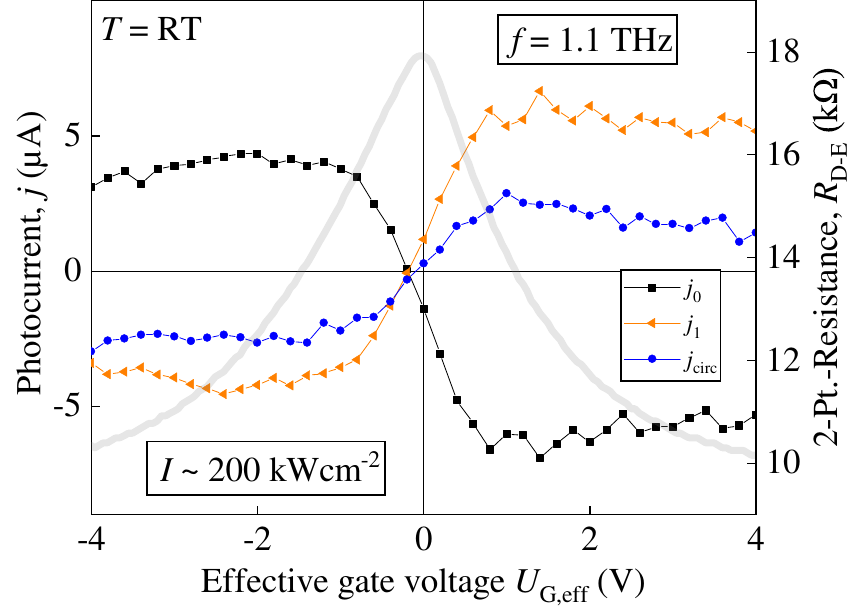}
	\caption{Dependences of the photocurrent contributions $j_{0}$, $j_{1}$, and $j_{\text{circ}}$ (see Eq.~\eqref{parameters_definition} for the $y$-direction (see inset in Fig.~\ref{Fig1}) on the effective back gate voltage $U_{\text{G,eff}}$ for 
	room temperature. Thick gray lines show the corresponding two-point resistance $R_{\text{D-E}}$.}
	\label{Fig3}
\end{figure}

Measuring the photocurrent and the two-point resistance at room temperature we observed that the oscillations with the gate voltage disappear. The photocurrent is almost constant for high gate voltages and changes its sign at the \gls{cnp}. This is shown for different photocurrent contributions in Fig.~\ref{Fig3}. 
Note that the contributions $j_0$ and $j_1$ were extracted by measuring the dependence of the photocurrent on the gate voltage for $\alpha=\Degree{0}$ and $\alpha=\Degree{90}$, and calculating the half-sum ($j_0$) or half-difference ($j_1$) of the corresponding photocurrents, respectively. Similarly, the circular contribution was extracted by taking the half-difference of photocurrents excited by right- and left-handed circularly polarized radiation.
Studying the intensity dependence for room temperature, we observed that the photocurrent saturates, whereas the saturation intensity increases strongly (not shown).

\begin{figure}
	\centering
	\includegraphics[width=\linewidth]{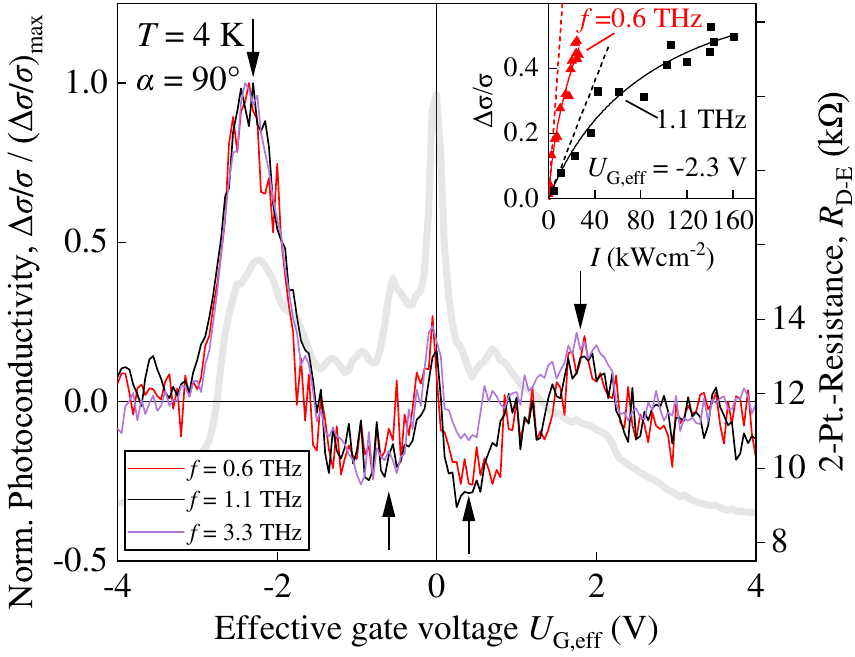}
	\caption{Dependences of the photoconductivity $\Delta\sigma/\sigma$ in the $y$-direction (see inset in Fig.~\ref{Fig1}) on the effective gate voltage $U_{\text{G,eff}}$ for different radiation frequencies. The curves are normalized to their maximum value. The thick gray line show the corresponding two-point resistance $R_{\text{D-E}}$. The corresponding radiation intensities are \kWcm{45} (\TeraHertz{0.6}), 
		\kWcm{200} (\TeraHertz{1.1}), and 
		\kWcm{380} (\TeraHertz{3.3})
		. The inset depicts the intensity dependences of the photoconductivity $\Delta\sigma/\sigma$ as a function of the radiation intensity for an effective gate voltage of \SI{-2.3}{\volt} and radiation frequencies of 0.6 and \TeraHertz{1.1}.}
	\label{Fig4}
\end{figure}

To obtain a complete picture of the observed nonlinearity, we studied the radiation-induced change of conductivity $\Delta\sigma/\sigma$. Figure~\ref{Fig4} shows the gate voltage dependence of the photoconductivity measured for different frequencies. As addressed above the photoconductivity is extracted from the response by taking the half-difference between photosignals obtained for positive and negative bias voltages. In order to simplify the comparison between the photoresponse at different frequencies in Fig.~\ref{Fig4}, we normalized the photoconductivity on its maximum values obtained at $U_{\text{G,eff}}= \SI{-2.3}{\volt}$. Note that the maximum of the photoconductivity coincide with an oscillation maximum of the sample resistance, see Fig.~\ref{Fig4}. The photoconductivity signal exhibits pronounced sign-alternating oscillations with the gate voltage, which, alike the photocurrent, correlates with the resistance. Importantly, the oscillations are detected even at highest intensities and, because of the change of sign, are pronounced even better. Furthermore, in contrast to the photocurrent, the photoconductive signal doesn't change its sign in the vicinity of the \gls{cnp}. Note that the photoconductivity signal shows only a slight dependence on the orientation of the radiation electric field $\bm E$ (not shown). Studying the intensity dependence of the photoconductivity we observed that for all gate voltages it saturates with the increase of intensity. The intensity dependence is exemplary shown in the inset in Fig.~\ref{Fig4} for an effective gate voltage of \SI{-2.3}{\volt} at radiation frequencies of 0.6 and \TeraHertz{1.1}. 

\begin{figure}
	\centering
	\includegraphics[width=\linewidth]{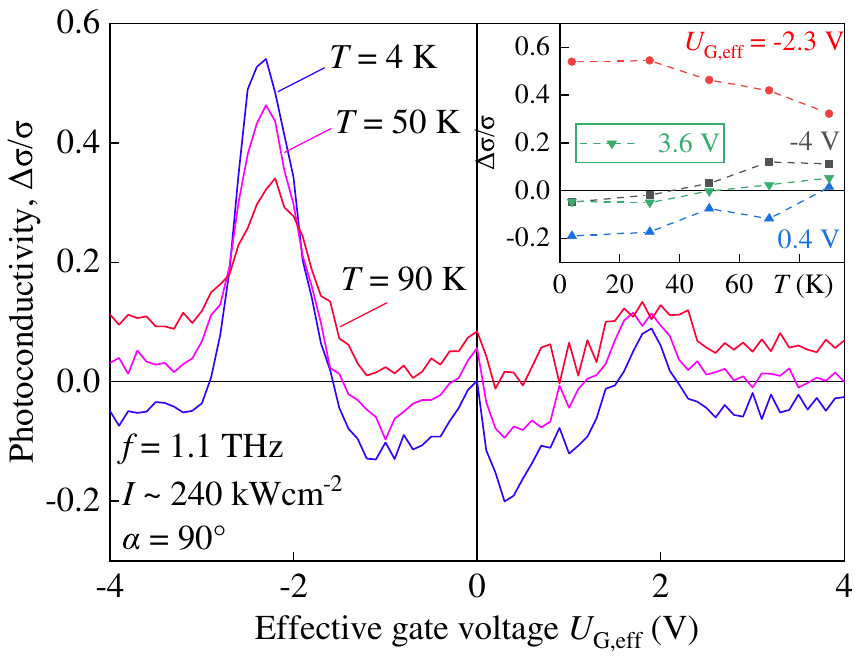}
	\caption{Dependences of the photoconductivity $\Delta\sigma/\sigma$ in the $y$-direction (see inset in Fig.~\ref{Fig1}) on the effective gate voltage $U_{\text{G,eff}}$ for different temperatures at a radiation frequency of \TeraHertz{1.1}. The corresponding radiation intensity is \kWcm{240}. The inset shows the dependences of the photoconductivity at various gate voltages as a function of the temperature.}
	\label{Fig5}
\end{figure}

An increase of the sample temperature results in a qualitative change of the effective gate voltage dependence, see Fig.~\ref{Fig5}. At high temperature ($T\geq\SI{90}{\kelvin}$) the formerly observed changes of photoconductivity sign are not present, whereas the oscillations are still clearly detected. The inset in Fig.~\ref{Fig5} shows the dependence on the photoconductivity measured at  $U_{\text{G,eff}}=\SI{-2.3}{\volt}$ at which the photoconductivity approaches its maximum, and three traces at different gate voltages as examples of the temperature-dependent sign inversion.

\section{Discussion}
\label{discussion}

\subsection{Nonlinear photogalvanic current}
%
%
%

As addressed above the observed photocurrent described by Eq.~(\ref{parameters_definition}) is caused by the bulk photogalvanic effect. The microscopic theory of this photocurrent is presented in Ref.~\cite{Otteneder2020}. It reveals that the three photocurrent contributions, which are given by the coefficients $j_0, j_1$, $j_2$, and the corresponding Stokes parameters, are caused by the electron scattering asymmetry of the photoexcited carriers and are proportional to the radiation absorbance $\eta$. The latter determines the variation of the photocurrent with the radiation intensity, see e.g. Ref.~\cite{Ganichev2002,Candussio2021a,Danilov2021}. In the considered twisted graphene structures, absorption of \gls{thz} radiation with photon energies of the order of several \si{\milli\electronvolt} can, in general, be caused by two mechanisms: The free-carrier absorption due to indirect \emph{intra}band optical transitions (Drude-like), and  the direct \emph{inter}band/\emph{intersub}band optical transitions \footnote{Note that in the vicinity of \gls{cnp} we obtain direct optical transitions between valence and conduction bands, which consequently result in the generation of electron-hole pairs. For higher gate voltages, however, direct optical transitions appear between Moiré subbands and result in a photoexcitation of free carriers.}.

Figure~\ref{Fig6}(a) shows that for most gate voltages the photocurrent saturates with increasing radiation intensity. For the studied range of radiation intensities absorption bleaching is indeed the most favorable nonlinear process for both absorption mechanisms~\cite{Ganichev2002,Candussio2021a,Danilov2021}, because other effects, which may cause the nonlinear absorption,  such as e.g. the two-photon absorption, require much higher intensities to yield a contribution comparable with the one-photon absorption channel~\cite{Candussio2021a}. The bleaching of the Drude-like radiation absorption in monolayer graphene has recently been observed by means of nonlinear ultrafast  \cite{Mics2015} and nonlinear photogalvanic \gls{thz} spectroscopy~\cite{Candussio2021a}. The range of radiation frequencies (0.4–\TeraHertz{1.2}) and electric fields  (2 – \kVcm{100})  used in these studies are similar to that used in our work. It has been shown that the absorption bleaching is caused by electron gas heating followed by energy relaxation and is well described by an empirical analytical formula 
\begin{equation}	
\eta^{\rm Dr} \propto \left(1+ \frac I{I_{\text s}^{\rm Dr}}\right)^{-1}\,, \label{eq_absorbance1}
\end{equation}
with the saturation intensity $I_{\text{s}}^{\rm Dr}$  being proportional to the reciprocal energy relaxation time and the Drude absorption cross-section. Absorption and photogalvanic current saturation for direct optical transitions have been theoretically and experimentally studied in monolayer graphene, see Ref.~\cite{Candussio2021a}, respectively. The theory developed for the Dirac energy dispersion yields that at high intensities the absorption varies with the radiation intensity as~\cite{Leppenen2021, Leppenen2019} 
	%
	\begin{equation}		
	\eta^{\text{ib}} \propto \left(1+\frac{I}{I_{\text{s}}^{\text{ib}}}\right)^{-1/2}\,, \label{eq_absorbance2}
\end{equation}		
where $I_{\text{s}}^{\text{ib}}$ is the corresponding saturation intensity. Such functional behavior is also known for the inhomogeneous model of direct optical transitions~\cite{Demtroeder2008,Ganichev2005}. For the complex band structure of \gls{tblg} with a twist angle close to the magical one, these theories cannot be directly applied and the required analytical equation is not developed as yet. Nevertheless, as we show below,  Eq.~\ref{eq_absorbance2}, together with Eq.~\ref{eq_absorbance1}, describes well all experimental findings.

We begin with the polarization-independent contribution $j_0$, which yields the dominant contribution to the bulk photogalvanic current, see Fig.~\ref{Fig1}. We assume that the total photocurrent is caused by an interplay of the photogalvanic current related to the absorption due to indirect and direct optical transitions, whose nonlinear behavior is described by the equations above. Consequently, the total photocurrent is given by
	\begin{equation}
		j_{0}=A^{\text{Dr}} I \left(1+ \frac I{I_s^{\rm Dr}}\right)^{-1}+A^{\text{ib}} I \left(1+\frac{I}{I_{\text{s}}^{\text{ib}}}\right)^{-1/2}\,,
		\label{eq_photocurrent_fit}
	\end{equation}
with fitting parameters $A^{\text{Dr}}$, $I_{\text{s}}^{\text{Dr}}$, $A^{\text{ib}}$, and $I_{\text{s}}^{\text{ib}}$. Here $A^{\text{Dr}} $ and  $A^{\text{ib}}$ are low-intensity amplitudes of the PGE caused by indirect and direct optical transitions, respectively. Corresponding fits are shown in Fig.~\ref{Fig6} demonstrating that the equation describes well the experimental data. Fitting parameters used for calculation of these curves are given in Tab.~\ref{tab1}. The nonlinearity of the linear and circular photogalvanic effects given respectively by the coefficients $j_1$ and $j_c$ are also well described by the Eq.~\eqref{eq_photocurrent_fit} (not shown) \footnote{Note that the second contribution of the linear photogalvanic effect given by the coefficient $j_2$ is almost zero, which excludes study of its nonlinearity.}. Therefore, to be specific, below we focus our discussion on the nonlinearity of the polarization-independent contribution $j_0$.

Fitting the data for different gate voltages we observed that all fitting parameters vary non-monotonically, see Tab.~\ref{tab1}. First, we discuss the amplitudes of the photocurrents excited due to indirect and direct optical transitions. Table~\ref{tab1} shows that the amplitude $A^{\text{Dr}}$ of the photocurrent caused by the Drude absorption changes the sign by switching from positive to negative gate voltages, i.e. it has opposite polarity for electron and hole conductivity.  This is a direct consequence of the PGE current, which in the absence of an external magnetic field is described by an odd function of the carrier charge~\cite{Candussio2021a}. Microscopically, this photocurrent is generated due to asymmetric scattering of electrons driven back and forth by the \gls{thz} field. Both radiation absorption and scattering asymmetry depend on the details of the band structure. In the studied twisted graphene the \gls{thz} radiation-induced photocurrent is formed due to indirect optical transitions in one of the Moiré bands. Consequently, for different gate voltage, i.e. different Fermi level position, it is generated in different Moiré bands. Calculations of the band structure show that these bands are characterized by very different energy dispersions~\cite{Morell2010,Choi2019,Otteneder2020,Lu2021,Wu2021}, and therefore it is not surprising that the coefficients  $A^{\text{Dr}}$ for various gate voltages differs by several times.

For the direct optical transitions, the highest values of the coefficients $A^{\text{ib}} > 0 $ are obtained for the gate voltages corresponding to the  CNP vicinity, see Tab.~\ref{tab1}. This photocurrent is attributed to the direct interband transitions resulting in the generation of the electron-hole pairs followed by asymmetric scattering. Its amplitude depends on the difference in occupation of the initial and final states of the optical transitions.   The sign of this current is  defined only by the difference of the electron and hole momentum relaxation times and does not change by switching from electron to hole conductivity. Note that for the positive gate voltages the amplitude $A^{\text{ib}}$ is positive, whereas  $A^{\text{Dr}}$ is negative, and therefore for the photocurrent extreme at $U_{\text{G,eff}}=\SI 1{\volt}$, see vertical arrow in Fig.~\ref{Fig2}(a), these current almost completely compensate each other, so that the total current becomes zero. The photocurrent due to the direct interband transitions is generated in the vicinity of the CNP and vanishes at higher positive/negative gate voltages. At large negative gate voltages (about -2.5~V), however, we observed that the direct optical transitions yield a considerable contribution to the total photocurrent, see Tab.~\ref{tab1}. 
At these $U_{\text{G,eff}}$ the Fermi level is located in the gap between two  Moiré subbands, which corresponds to the highest difference in occupations of initial and final states, and enables direct optical transitions between these bands. Consequently, we obtain the peak of the photocurrent and the maximum of the sample resistance, see vertical arrow in Fig.~\ref{Fig2}(a). 
Outside the peak amplitudes, $A^{\text{ib}}$ decreases and approaches zero.

The most clear evidence for the presence of two competing mechanisms of the photocurrent, which are caused by indirect and direct optical transitions characterized by different saturation intensities, comes from the results obtained for low positive gate voltages. Indeed, in
most of the cases we observed that for negative gate voltages the total photocurrent $j_0$ has a positive sign in the whole range of used radiation intensities, and for positive gate voltages its sign reverses, see Fig.~\ref{Fig6}(a). For low positive gate voltages, however, we detected that an increase of the radiation intensity results in the photocurrent sign inversion, see Fig.~\ref{Fig6}(b). 
For low gate voltages, being close to \gls{cnp}, the contribution of the indirect optical transitions to the photocurrent is small, because these voltages are close to the sign inversion at the \gls{cnp} at which $A^{\text{Dr}}$ vanishes due to the sign inversion. Consequently, the photocurrent is dominated by the interband mechanism ($ A^{\text{ib}} > |A^{\text{Dr}}| $), which results in a positive photocurrent sign, see Tab.~\ref{tab1}. An increase of the radiation intensity, however, results in the saturation of this mechanism with $I_{\text{s}}^{\text{ib}} < I_{\text{s}}^{\text{Dr}}$,  see Tab.~\ref{tab1}, and the Drude contribution to the photocurrent shows up. The latter one has a negative sign and, respectively, the sign of the total photocurrent changes. Fitting functions in Fig.~\ref{Fig6}(b) show that the above scenario indeed describes well the complex intensity behavior observed for low positive gate voltages.

Now we discuss the variation of the saturation intensities with the gate voltage. As addressed above, both $I_s^{\rm Dr}$ and $I_{\text{s}}^{\text{ib}}$ vary non-monotonically with $U_{\text{G,eff}}$, see Tab.~\ref{tab1}. In general, the value of the saturation intensities is defined by the absorption cross-section and relaxation times. In \gls{tblg} with small twist angles the band structure is very complex. It consists of multiple flat Moiré subbands and variation of the Fermi level position makes possible both the  optical transitions within different subbands  (indirect transitions) as well as between them (direct transitions). The carriers in these bands have different masses and are characterized by different relaxation times. Changes of these parameters modify both, the absorption cross-section and relaxation times, which affect the saturation intensities. Furthermore, probabilities of direct interband/intersubband transitions are proportional to the difference in occupations of the corresponding initial and final states, which depends on the Fermi energy position, i.e., the effective gate voltage. These arguments explain why the saturation intensities for different gate voltages vary strongly (from several to tens \si{\kilo\watt\per\centi\meter\squared}), but not monotonically.

For negative gate voltages, for which both amplitudes $A^{\text{ib}}$ and $A^{\text{Dr}}$ have the same sign, and for large positive gate voltages at which the Drude PGE dominates, the sign inversion with increase of the radiation intensity is not detected and the photocurrent saturates with saturation intensities depending on the gate voltage.

The data discussed above are obtained at $T=\SI{4}{\kelvin}$.  Measurements for higher temperatures performed for several characteristic gate voltages demonstrated that up to about $T= \SI{100}{\kelvin}$ the photocurrent magnitude and its nonlinearity does not change much, see Fig.~\ref{Fig8}. This shows that the Drude absorption, being sensitive to the carrier scattering time, remains almost the same in this temperature range. Indeed, in transport measurements we observed that the sample resistance and, consequently, the momentum scattering time doesn't change much in this temperature range. At room temperatures, however, the overall behavior of the PGE changes substantially. First of all, no oscillations of the photocurrent upon variation of  the gate voltages can be resolved, see Fig.~\ref{Fig3}. This is not surprising, considering that the forbidden gaps and used photon energies are of the order of several \si{\milli\electronvolt}, whereas for room temperature $k_{\text B}T$ is substantially higher being about $\SI{26}{\milli\electronvolt}$, here $k_{\text B}$ is the Boltzmann constant. Furthermore, the saturation intensities at room temperature are observed to be substantially higher than those at low temperatures. The increase of the saturation intensities is caused by the enhancement of the scattering by phonons at room temperature, which substantially reduces the relaxation times.


\subsection{Nonlinear photoconductivity}

Finally, we briefly discuss the observed photoconductivity. Alike the photogalvanic effects discussed in previous section the photoconductive response is related to different absorption mechanisms caused by indirect-, direct intersubband  and direct interband optical transitions. In two former cases (Drude-like and intersubband absorption), the number of free carriers does not change and the radiation induced change of the sample conductance is caused by the electron gas heating and the associated change of the carrier mobility. For the Drude absorption the photoconductivity is described by $\Delta \sigma = e n \Delta \mu_n$ for electrons and $\Delta \sigma = e p \Delta \mu_p$, where $\mu_n$ and $\mu_p$ are electron and hole mobilities in the  Moiré subbands. For scattering by acoustic phonons the heating results in the negative photoconductivity, i.e. the conductivity decreases.  In contrast to the photocurrent, the sign of the photoconductivity does not depend on the type of carriers. At direct intersubband optical transitions THz radiation excites free carriers in the upper lying Moiré subband. This process also results in a change of the conductivity, because different Moiré subbands may have different mobilities.  On the other hand, absorption of radiation results in the electron gas heating, and, as addressed above, in the change of $\mu_n$ or $\mu_p$. In the  case of the direct optical transitions between the valence and conduction bands, THz radiation results in the generation of electron-hole pairs and, consequently, increases the sample conductance (positive photoconductivity). The opposite sign of the photoconductive response caused by the Drude- and interband-contributions explains the  sign inversions as a function of the gate voltage, which are observed at \SI{4}{\kelvin} in the vicinity of the \gls{cnp}, see Fig.~\ref{Fig4}. At the \gls{cnp}, at which the Fermi level lies in the forbidden gap, the interband transitions yield the positive photoconductivity which dominates the photoresponse. An increase of $|U_{\text{G,eff}}|$ decreases this contribution (the difference between the population of the initial and final states decreases) but increases the Drude absorption and, the magnitude of the corresponding negative photoconductivity. 
Consequently, the sign of the photoresponse reverses. At further increase of the gate voltages the photoresponse exhibits a resonance-like behavior and the sign reverses again, see Fig.~\ref{Fig4}. The resonances, marked in the figure by downward arrows, are caused by the direct intersubband transitions. Figure~\ref{Fig4} reveals that these transitions yield the positive photoconductivity (\gls{thz} radiation decreases the sample resistance).

While the oscillations of the photoconductance are clearly detected up to rather high temperatures ($T \approx \SI{100}{\kelvin}$), the increase of temperature results in the disappearance of the sign inversions, see Fig.~\ref{Fig5}. The temperature dependence of the photoconductivity measured at different gate voltages (at signal maximum, high positive/negative voltages and low positive gate voltages) shows that the lowest temperature inversion values are detected at high gate voltages  of both polarities ($U_{\text{G,eff}} < \SI{-3}{\volt}$ and $U_{\text{G,eff}} > \SI{2}{\volt}$). These conditions correspond to the Drude mechanism being the dominant mechanism for of the photoconductivity. Opposite signs of the photoconductive response detected for $T=4$ and $T>\SI{40}{\kelvin}$ we attribute to  the change of the temperature dependence of the carrier mobility and, consequently, the sign of $\Delta \mu_p$ and $\Delta \mu_n$. Note that the change of sign of $\partial R/\partial T$ is also detected in the transport measurements (not shown).	Understanding of the observed  temperature sign inversion requires further study and is out of scope of this paper.

Alike the photocurrent, the photoconductive signal saturates with an increase of the radiation intensity, see the inset in Fig.~\ref{Fig4}. Taking into account the nonlinearity of the corresponding absorbance (see Eqs.~\eqref{eq_absorbance1} and \eqref{eq_absorbance2}) the nonlinearity of the total photoconductivity is described by
\begin{equation}
\frac{\Delta\sigma}{\sigma}=B^{\text{Dr}}I\left(1+ \frac I{I_{\text s}^{\rm Dr}}\right)^{-1}+B^{\text{ib}}I\left(1+\frac{I}{I_{\text{s}}^{\text{ib}}}\right)^{-1/2},
\end{equation}
with fitting parameters $B^{\text{Dr}}$, $I_{\text{s}}^{\text{Dr}}$, $B^{\text{ib}}$, and $I_{\text{s}}^{\text{ib}}$. Here $B^{\text{Dr}}$ and  $B^{\text{ib}}$ are low-intensity amplitudes of the photoconductivity responses caused by indirect and direct optical transitions, respectively. For $T=\SI{4}{\kelvin}$ we observed that the amplitude  $B^{\text{Dr}}$  is negative and $B^{\text{ib}}$ is positive. The inset in Fig.~\ref{Fig4} shows that the above equation describes well the nonlinear behavior of the photoconductivity. The inset also shows that the decrease of the radiation frequency results in a substantial increase of the photoconductive signal. 
The increase of the absorption cross-section with the frequency reduction should result in the decrease of the saturation intensity, which is indeed detected in experiment, see the inset in Fig.~\ref{Fig4}.

\begin{table*}[]
	\centering
	\begin{tabular}{c|c|c|c|c|c}
		\makecell{Eff. Gate Voltage,\\$U_{\text{G,eff}}$ (V)} & \makecell{Frequency,\\ $f$(\si{\tera\hertz})}& \makecell{$A^{\text{Dr}}$ \\(\si{\micro\ampere \centi\meter\squared \per\kilo\watt})}& \makecell{$I_{\text{s}}^{\text{Dr}}$\\ (\si{\kilo\watt\per\centi\meter\squared})} & \makecell{$A^{\text{ib}}$\\ (\si{\micro\ampere \centi\meter\squared \per\kilo\watt})}&\makecell{ $I_{\text{s}}^{\text{ib}}$\\ (\si{\kilo\watt\per\centi\meter\squared})} \\ \hline
		\hline 
		\rule{0pt}{3ex}
		-2.8 & 0.6 & 3.30  & 1.0   & 0.20 & 10.0  \\
		-2.2  & 0.6 & 0.36  & 5.0   & 0.18 & 25.0  \\
		-1.6 & 0.6 & 0.15  & 115.0 & 0 & - \\
		-0.4 & 0.6 & 0.14  & 5.0   & 0.38 & 1.5  \\
		0.6  & 0.6 & -0.67 & 64.0  & 0.87 & 7.6   \\
		1.0  & 0.6 & -0.40 & 88.0 & 0.60 & 6.0 \\
		2.4 & 0.6 & -0.14 & 79.6  & 0 & - \\
		3.4 & 0.6 & -0.10 & 65.0 & 0 & - \\
		\hline
		\rule{0pt}{2.5ex}
		-2.5 & 1.1 & 0.56 & 2.5 & 0.18 & 1.4\\
		2.5 & 1.1 & -0.05 & 122.0 & 0 & - \\
		-1.0 & 1.1 & 0.19 & 6.0 & 0.04 & 32.4 \\
	\end{tabular}
\caption{Values of the parameters used for the fits of the data at $T = \SI{4}{\kelvin}$ presented in Figs.~\ref{Fig6} and \ref{Fig8}.}
\label{tab1}
\end{table*}

\section{Summary} 
\label{summary}

To summarize, the observed nonlinear intensity dependence of the \gls{thz} radiation-induced bulk photogalvanic current can be well described by the interplay of two microscopic mechanisms related to the Drude-like free carrier absorption and direct interband/intersubband  optical transitions. Both types of the photocurrent saturate at high intensities and are characterized by different values of their magnitudes and saturation intensities. While the photocurrent related to the free carrier absorption is generated at all gate voltages apart the \gls{cnp} at which it changes the sign, the photocurrent related to the direct optical transitions is only present for specific gate voltages  corresponding to the maxima of the sample resistance. The observed saturation of the Drude contribution is attributed to the electron gas heating, whereas the one excited by the direct optical 
transitions is caused by the slow energy relaxation of the photoexcited carriers.
Due to the complex band structures the magnitudes and the saturation intensities of the photocurrent varies non-monotonically with the gate voltage. The excitation with intense terahertz radiation also results in the change of the \gls{tblg} conductivity which is related to different absorption channels. On the one hand, the photoconductivity due to free carrier absorption is caused by the electron gas heating resulting in the change of the carrier mobility. On the other hand,  the photoconductive response is excited due to direct interband transitions and direct optical transitions between Moiré subband. The direct optical transitions result in the resonance-like behavior of the photoconductivity as a function of the gate voltage. The interplay of all three  contributions results in sign-alternating oscillations of the photoresponse with extrema positions corresponding to that of the sample resistance and the photogalvanic current. Alike the photogalvanic current,  the photoconductive response saturates at high intensities, which is also caused by the nonlinearity of the radiation absorption. The observed saturation of the photocurrent and photoconductive response provide an access to study of scattering processes in twisted graphene with small twist angles close to the magical one. 

\section{Acknowledgments}
\label{acknow} 
We thank V.V. Bel'kov and L.E. Golub for fruitful discussions. The support from the Deutsche Forschungsgemeinschaft (DFG, German Research Foundation) - Project SPP 2244 (GA501/17-1), and the Volkswagen Stiftung Program (97738) is gratefully acknowledged. S.D.G. thanks the support from the IRAP program of the Foundation for Polish Science (grant MAB/2018/9, project CENTERA). D.K.E. acknowledges support from the Ministry of Economy and Competitiveness of Spain through the ‘Severo Ochoa’ programme for Centres of Excellence in R and D (SE5-0522), Fundacio Privada Cellex, Fundacio Privada Mir-Puig, the Generalitat de Catalunya throughthe CERCA programme, funding from the European Research Council (ERC) under the European Union’s Horizon 2020 research and innovation programme (grant agreement no. 852927). 
K.W. and T.T. acknowledge support from the Elemental Strategy Initiative conducted by the MEXT, Japan (Grant Number JPMXP0112101001) and  JSPS KAKENHI (Grant Numbers 19H05790, 20H00354 and 21H05233).


\appendix

\IfFileExists{I:/_Papers_Abstracts/JabRef_Bibliography/all_lib.bib}{\bibliography{I:/_Papers_Abstracts/JabRef_Bibliography/all_lib.bib}}{\bibliography{all_lib}}

\end{document}